\documentclass[epj]{svjour}
\usepackage{latexsym}
\usepackage{graphics}
\usepackage{amssymb}
\usepackage{amsmath,bm}
\usepackage{graphicx}
\usepackage[normalem]{ulem}
\usepackage[dvips]{color}
\renewcommand\sout{\bgroup \color{red} \ULdepth=-.5ex \ULset}

\begin{document}
\title{Relationship between the symmetry energy and the single-nucleon potential in isospin-asymmetric nucleonic matter}
\author{Chang Xu\inst{1}, Bao-An Li\inst{2}, \and Lie-Wen Chen\inst{3}
}                     
\offprints{}          
\institute{School of Physics, Nanjing University, Nanjing 210008,
China \and Department of Physics and Astronomy, Texas A$\&$M
University-Commerce, Commerce, Texas 75429-3011, USA \and
Department of Physics and Astronomy and Shanghai Key Laboratory
for Particle Physics and Cosmology, Shanghai Jiao Tong University,
Shanghai 200240, China}
\date{Received: date / Revised version: date}
%
\abstract{In this contribution, we review the most important
physics presented originally in our recent publications
\cite{Xuli10b,xulinpa,Xuli10a,xuli,Che12a,Che12b}. Some new
analyses, insights and perspectives are also provided. We showed
recently that the symmetry energy $E_{sym}(\rho )$ and its density
slope $L(\rho)$ at an arbitrary density $\rho$ can be expressed
analytically in terms of the magnitude and momentum dependence of
the single-nucleon potentials by using the Hugenholtz-Van Hove
(HVH) theorem. These relationships provide new insights about the
fundamental physics governing the density dependence of nuclear
symmetry energy. Using the isospin and momentum (k) dependent MDI
interaction as an example, the contribution of different terms in
the single-nucleon potential to the $E_{sym}(\rho )$ and $L(\rho
)$ are analyzed in detail at different densities. It is shown that
the behavior of $E_{sym}(\rho )$ is mainly determined by the
first-order symmetry potential $U_{sym,1}(\rho ,k)$ of the
single-nucleon potential. The density slope $L(\rho )$ depends not
only on the first-order symmetry potential $U_{sym,1}(\rho ,k)$
but also the second-order one $U_{sym,2}(\rho ,k)$. Both the
$U_{sym,1}(\rho ,k)$ and $U_{sym,2}(\rho ,k)$ at normal density
$\rho_0$ are constrained by the isospin and momentum dependent
nucleon optical potential extracted from the available
nucleon-nucleus scattering data. The $U_{sym,2}(\rho ,k)$
especially at high density and momentum affects significantly the
$L(\rho)$, but it is theoretically poorly understood and currently
there is almost no experimental constraints known.  
} 
\maketitle
\section{Introduction}
In recent years, extensive experimental and theoretical efforts
have been devoted to determining the density dependence of nuclear
symmetry energy $E_{sym}(\rho)$, which characterizes the isospin
dependent part of the equation of state (EOS) of asymmetric
nuclear matter \cite{li0,li1,li2}. Starting from a model energy
density functional, the symmetry energy $E_{sym}(\rho)$ and its
density slope $L(\rho)=3\rho\frac{\partial E_{sym}(\rho)}{\partial
\rho}$ can be easily obtained by expanding the EOS of asymmetric
nuclear matter as a power series of isospin asymmetry
$\delta=\frac{\rho_n-\rho_p}{\rho_n+\rho_p}$: $E(\rho ,\delta
)=E_{0}(\rho )+E_{\mathrm{sym}}(\rho )\delta ^{2}+O(\delta ^{4})$
where $E_{0}(\rho )$ is the EOS of symmetric nuclear matter and
$E_{\mathrm{sym}}(\rho )$ is the so-called symmetry energy.
Although much information about the EOS of symmetric nuclear
matter $E_0(\rho)$ has been accumulated over the past several
decades, our knowledge about the $E_{sym}(\rho)$ is unfortunately
still very poor. However, it has been clearly shown in many
references that the symmetry energy and its density slope are
critical for understanding not only the structure of rare isotopes
and the reaction mechanism of heavy-ion reactions, but also many
interesting issues in astrophysics
\cite{JML04,AWS05,dan,sie70,lee98,ste,sjo,bro,bar,Toro10,Sum94,Bom01}.
Thus, to determine the density dependence of $E_{sym}(\rho)$ has
now become a major goal in both nuclear physics and astrophysics.
While significant progress has been made recently in constraining
the $E_{sym}(\rho)$ especially below and around the normal
density, see, e.g., \cite{LWC05a,LWC05b,tsa,Cen09,Joe10}, much
more work needs to be done to constrain the $E_{sym}(\rho)$ at
high densities
\cite{xia,Das03a,Das03b,Das03c,ulr,van,zuo,Fri05,she,Che05c,zhli,Sto03,pan,wir,kut}.
The present theoretical predictions on the high density behavior
of $E_{sym}(\rho)$ are rather diverse by various
nonrelativistic/relativistic mean-field approaches, depending
closely on the mean-field/single-nucleon potential used in the
model.

As an important input for calculations of nuclear structures and
simulations of heavy-ion reactions, the single-nucleon potential
$U_{n/p}(\rho,\delta,p)$ itself can also be obtained by a
functional derivative of the energy density $\xi=\rho E(\rho
,\delta )$ with respect to the distribution function
\cite{Das03a,Das03b,Das03c}. Thus, the single-nucleon potential
and the symmetry energy are intrinsically correlated as they can
be both obtained from the same energy density functional. In this
paper, we will review the direct relationship between the symmetry
energy and the single-nucleon potential $U_{n/p}(\rho,\delta,p)$
\cite{Xuli10b,xulinpa,Xuli10a,xuli,Che12a,Che12b}. For studying
the symmetry energy and its density slope, the direct relationship
between the single-nucleon potential and the symmetry energy
without going through the procedure to construct the corresponding
energy density functional is obviously advantageous. This is
because one can directly extract both the isoscalar and isovector
nucleon optical potentials at saturation density from experimental
data, such as (p,n) charge exchange reactions and
proton/neutron-nucleus scattering. One can then easily calculate
the symmetry energy and its density slope at saturation density
directly from the optical potentials without having to first
construct the energy density functional \cite{Xuli10b}. Moreover,
to find the relationship between the symmetry energy and the
isoscalar and isovector single-nucleon potentials is actually a
major goal of the current efforts in developing nuclear energy
density functionals \cite{INT2005}. It is also mentioned that,
within relativistic covariant formulism, the Lorentz covariant
nucleon self-energy decomposition of the nuclear symmetry energy
has also been obtained recently~\cite{CaiBJ12}.

In this paper, we shall firstly recall the general relationship
between the symmetry energy and the single-nucleon potential in
isospin asymmetric matter derived earlier in Refs.
\cite{Xuli10b,xulinpa,Che12a} using the Hugenholtz-Van Hove
theorem \cite{hug}. The analytical expressions of the symmetry
energy $E_{sym}(\rho)$ and its density slope $L(\rho )$ are very
helpful in gaining deeper insights into the microscopic origins of
$E_{sym}(\rho)$ and $L(\rho)$. Using the isospin and momentum
dependent MDI interaction as an example
\cite{Das03a,Das03b,Das03c}, the contributions of different terms
in the single-nucleon potential (MDI) to the $E_{sym}(\rho )$ and
$L(\rho )$ are analyzed in details for different densities. The
outline of this paper is as follows. In Section 2, the
relationship between the symmetry energy and the single-nucleon
potential is derived by using the Hugenholtz-Van Hove theorem. In
Section 3, the isoscalar and isovector potentials of the isospin
and momentum dependent MDI interaction are introduced in details.
The first-order and second-order symmetry potentials of the MDI
interaction are also compared with those from several microscopic
approaches in Section 3. The optical model analysis of the
single-nucleon potential and the corresponding symmetry energy and
its density slope are presented in Section 4. Finally a brief
summary is given in Section 5.

\section{Relationship between the symmetry energy and the single-nucleon potential based on the Hugenholtz-Van Hove theorem}
In 1958, Hugenholtz and Van Hove proposed a famous theorem on the
single particle energy in a Fermi gas with interaction at absolute
zero in temperature \cite{hug}. In the following, we recall the
Hugenholtz-Van Hove theorem and its application in deriving the
relation between the nuclear symmetry energy and the
single-nucleon potential. The HVH theorem describes a fundamental
relation among the Fermi energy $E_{F}$, the average energy per
particle $E$ and the pressure of the system $P$ at zero
temperature. For a one-component system, in terms of the energy
density $\xi=\rho E$, the general HVH theorem can be written as
\cite{hug,Sat99}
\begin{eqnarray}\label{HVH}
E_{F}=\frac{d\xi}{d \rho} = E+ \frac{P}{\rho}.
\end{eqnarray}
For a special system with zero pressure the Fermi energy $E_{F}$
is equal to the average energy per particle $E$ of the system.
\begin{eqnarray}
E_{F}=E.
\end{eqnarray}
It is stressed that the general HVH theorem of Eq.(\ref{HVH}) is
valid at arbitrary density as long as the temperature $T$ remains
zero \cite{hug,Sat99}. It does not depend on the precise nature of
the interaction. In fact, a successful theory of nuclear matter is
required not only to describe properly the saturation properties
of nuclear matter but also to fulfill the HVH theorem at any
density. For instance, in the original paper of the HVH theorem
\cite{hug}, Hugenholtz and Van Hove used their theorem to test the
internal consistency of the nuclear matter theory of Brueckner.
They found the large discrepancy between the values of $E_{F}$ and
E in the Brueckner's theory at equilibrium and pointed out that
Brueckner neglected important cluster terms contributing to the
single particle energy \cite{hug}.

According to the HVH theorem, the Fermi energies of neutrons and
protons in isospin asymmetric nuclear matter are, respectively
\cite{xulinpa,Xuli10a,xuli,hug,Sat99,bru64,Dab73a,Dab73b,Dab73c},
\begin{eqnarray} t(k_F^n)+U_n(\rho,\delta,k_F^n) =
\frac{\partial \xi }{\partial \rho_n}, \label{chemUn}
\\
t(k_F^p)+U_p(\rho,\delta,k_F^p) = \frac{\partial \xi }{\partial
\rho_p}, \label{chemUp}
\end{eqnarray}
where $t(k)=\hbar k^2/2m$ is the kinetic energy and $U_{n/p}$ is
the neutron/proton single-nucleon potential. The Fermi momenta of
neutrons and protons are $k_F^n=k_F(1+\delta)^{1/3}$ and
$k_F^p=k_F(1-\delta)^{1/3}$, respectively.  Subtracting
Eq.(\ref{chemUp}) from Eq.(\ref{chemUn}) gives
\cite{bru64,Dab73a,Dab73b,Dab73c}
\begin{eqnarray}\label{UnminusUp}
&&[t(k_F^n)-t(k_F^p)]+[U_n(\rho,\delta,k_F^n)-U_p(\rho,\delta,k_F^p)]
\nonumber \\ && = \frac{\partial \xi }{\partial
\rho_n}-\frac{\partial \xi }{\partial \rho_p}.
\end{eqnarray}
The single-nucleon potentials can be expanded as a power series of
isospin asymmetry $\delta$ while respecting the charge symmetry of
nuclear interactions under the exchange of protons and neutrons,
\begin{eqnarray}\label{UnUp1}
&& U_{\tau}(\rho,\delta,k)\, = \,U_0(\rho,k) +
\sum_{i=1,2,3...}U_{sym,i}(\rho,k) ({\tau}\delta)^i  \\ \nonumber
&& = U_0(\rho,k) + U_{sym,1}(\rho,k) ({\tau} \delta) +U_{sym,2}(k)
({\tau} \delta)^2 +...
\end{eqnarray}
where $\tau$=1 $(-1)$ for neutrons (protons). If one neglects the
higher-order terms ($\delta^2$, $\delta^3$,...), Eq.(\ref{UnUp1})
reduces to the so-called Lane potential \cite{Lan62}. Expanding
both the kinetic and potential energies around the Fermi momentum
$k_F$, the left side of Eq.(\ref{UnminusUp}) can be further
written as
\begin{eqnarray}\label{leftside}
&&[t(k_F^n)\,-\,t(k_F^p)]+[U_n(\rho,\delta,k_F^n)\,-\,U_p(\rho,\delta,k_F^p)] \nonumber\\
=&&\sum_{i=1,2,3...}\frac{1}{i!}\frac{\partial^i [t(k)+U_0(\rho,k)]}{\partial k^i}|_{k_F} k_F^i \nonumber\\
\times&&[(\sum\limits_{j=1,2,3..}F(j)\delta^j)^i-(\sum\limits_{j=1,2,3..}F(j)(-\delta)^j)^i] \nonumber\\
+ && \sum_{l=1,2,3...}U_{sym,l}(\rho,k_F)[\delta^{l}-(-\delta)^{l}] \nonumber\\
+
&&\sum_{l=1,2,3...}\sum_{i=1,2,3...}\frac{1}{i!}\frac{\partial^iU_{sym,l}(\rho,k)}{\partial
k^i}|_{k_F} k_F^i
\nonumber \\
\times &&  [(\sum\limits_{j=1,2,3..}F(j)\delta^j)^i
\delta^{l}-(\sum\limits_{j=1,2,3..}F(j)(-\delta)^j)^i
(-\delta)^{l}]
\nonumber \\
= && [\frac{2}{3} \frac{\partial [t(k)+U_0(\rho,k)]}{\partial
k}|_{k_F}k_F + 2 U_{sym,1}(\rho,k_F)]\delta + ...,
\end{eqnarray}
where the function
$F(j)=\frac{1}{j!}[\frac{1}{3}(\frac{1}{3}-1)...(\frac{1}{3}-j+1)]$
is introduced. For the right side of Eq.(\ref{UnminusUp}),
expanding in powers of $\delta$ gives
\begin{eqnarray} \label{rightside}
  \frac{\partial \xi }{\partial \rho_n}-\frac{\partial \xi
}{\partial \rho_p}  =4E_{sym}(\rho )\delta +\mathcal{O}(\delta
^{3})
\end{eqnarray}
Comparing the coefficient of each $\delta^i$ term in
Eq.(\ref{leftside}) with that in Eq.(\ref{rightside}) then gives
the symmetry energy of any order. For instance, we derived the
most important quadratic term
\begin{eqnarray}\label{Esymexp}
E_{sym}(\rho) &=& \frac{1}{6} \frac{\partial
[t(k)+U_0(\rho,k)]}{\partial k}|_{k_F}k_F + \frac{1}{2}
U_{sym,1}(\rho,k_F)\nonumber \\ &=&\frac{1}{3} t(k_F) +
\frac{1}{6} \frac{\partial U_0}{\partial k}\mid _{k_F}\cdot k_F +
\frac{1}{2}U_{sym,1}(\rho,k_F)
\end{eqnarray}

By adding Eq.(\ref{chemUn}) and Eq.(\ref{chemUp}), the following
equation is obtained
\begin{eqnarray}\label{twonew}
&&
[t(k_F^n)+t(k_F^p)]+[U_n(\rho,\delta,k_F^n)+U_p(\rho,\delta,k_F^p)]
\nonumber \\ && =\frac{\partial \xi }{\partial
\rho_n}+\frac{\partial \xi }{\partial \rho_p}.
\end{eqnarray}
The right side of Eq.(\ref{twonew}) can be further written as
\begin{eqnarray}
\frac{\partial \xi }{\partial \rho_n}+\frac{\partial \xi
}{\partial \rho_p}&=&2E_{0}(\rho )+2\rho \frac{\partial E_{0}(\rho
)}{\partial \rho }  \nonumber
\\
&+&\Big[\frac{2}{3}L(\rho )-2E_{sym}(\rho )\Big]\delta
^{2}+\mathcal{O}(\delta ^{4}).
\end{eqnarray}
Expanding again both the kinetic and potential energies in the
left side of Eq.(\ref{twonew}) around $k_F$ and comparing the
corresponding coefficients of two sides in Eq.(\ref{twonew}), we
obtained the exact analytical equation of the density slope
$L(\rho)$
\begin{eqnarray}
L(\rho ) &=&
\frac{1}{6}%
\frac{\partial \lbrack t(k)+U_{0}(\rho ,k )]}{\partial
k}|_{k_{F}}\cdot k_{F} \nonumber \\
& + &\frac{1}{6}\frac{\partial ^{2}[t(k)+U_{0}(\rho ,k )]}{\partial k^{2}}%
|_{k_{F}}\cdot k_{F}^{2} + \frac{3}{2}U_{sym,1}(\rho
,k_{F})
\nonumber \\
 &+& \frac{\partial U_{sym,1}(\rho ,k )}{\partial
k}|_{k_{F}}\cdot k_{F}+3U_{sym,2}(\rho ,k_{F}). \label{Lexp}
\end{eqnarray}%
Similar to the HVH theorem, the analytical expressions in
Eq.(\ref{Esymexp}) and Eq.(\ref{Lexp}) are valid at any density.
The values of both $E_{sym}(\rho)$ and $L(\rho)$ can be easily
calculated simultaneously once the single-nucleon potential is
known. The most critical advantage of the expressions in
Eq.(\ref{Esymexp}) and Eq.(\ref{Lexp}) is that they allow us to
determine the $E_{sym}(\rho)$ and $L(\rho)$ directly from the
value and momentum dependence of the single-nucleon potential at
$\rho$. Essentially, this enables one to translate the task of
determining the density dependence of the symmetry energy into a
problem of finding the momentum dependence of the $U_0(\rho,k)$,
$U_{sym,1}(\rho,k)$, and $U_{sym,2}(\rho,k)$.

\section{The isoscalar and isovector potentials of the Momentum-Dependent-Interaction (MDI)}

Several famous single-nucleon potentials have been widely applied
in the transport model simulations for heavy-ion reactions.
Usually, these single-nucleon potentials are derived from their
corresponding energy density functional that has been carefully
adjusted to properties of nuclear matter. To show in details the
contribution of each term in Eq.(\ref{Esymexp}) and
Eq.(\ref{Lexp}) to the symmetry energy and its density slope, we
use the widely-used Momentum-Dependent-Interaction (MDI) as an
example \cite{Das03a,Das03b,Das03c}, which is derived from the
Hartree-Fock approximation using a modified Gogny effective
interaction \cite{dec}
\begin{eqnarray}\label{mdi}
&&U_{\tau}(\rho,\delta,\vec p) = A_u(x)\frac{\rho_{\tau'}}{\rho_0}
+A_l(x)\frac{\rho_{\tau}}{\rho_0}\nonumber\\
&&+B(\frac{\rho}{\rho_0})^{\sigma}(1-x\delta^2)-4\tau
x\frac{B}{\sigma+1}\frac{\rho^{\sigma-1}}{\rho_0^{\sigma}}\delta\rho_{\tau'}
\nonumber \\&& +\frac{2C_{\tau,\tau}}{\rho_0} \int
d^3p'\frac{f_{\tau}(\vec r,\vec p')}{1+(\vec p-\vec
p')^2/\Lambda^2} \nonumber \\&& +\frac{2C_{\tau,\tau'}}{\rho_0}
\int d^3p'\frac{f_{\tau'}(\vec r,\vec p')}{1+(\vec p-\vec
p')^2/\Lambda^2},
\end{eqnarray}
where $\tau=1$($-1$) for neutrons (protons) and $\tau\neq\tau'$;
$\sigma=4/3$ is the density-dependence parameter; $f_{\tau}(\vec
r,\vec p)$ is the phase space distribution function at coordinate
$\vec{r}$ and momentum $\vec{p}=\hbar\vec{k}$. The parameters $B,
C_{\tau,\tau}, C_{\tau,\tau'}$ and $\Lambda$ are fitted to the
nuclear matter saturation properties \cite{Das03a,Das03b,Das03c}.
The parameters $B$ and $\sigma$ in the MDI single-nucleon
potential are related to the $t_0$ and $\alpha$ in the Gogny
effective interaction via $t_0 = \frac{8}{3} \frac{B}{\sigma+1}
\frac{1}{\rho_0^{\sigma}}$ and $\sigma = \alpha + 1$ \cite{dec}.
The parameter $x$ has been introduced to vary the density
dependence of the symmetry energy while keep the properties of
symmetric nuclear matter unchanged~\cite{LWC05a,LWC05b} and it is
related to the spin (isospin)-dependence parameter $x_0$ via
$x=(1+2x_0)/3$ \cite{Xuli10a}. The momentum dependence of the
symmetry potential stems from the different strength parameters
$C_{\tau,\tau'}$ and $C_{\tau,\tau}$ for a nucleon of isospin
$\tau$ interacting with unlike and like nucleons. More
specifically, $C_{unlike}=-103.4$ MeV while $C_{like}=-11.7$ MeV.
The quantities $A_{u}(x)=-95.98-x\frac{2B}{\sigma +1}$ and
$A_{l}(x)=-120.57+x\frac{2B}{\sigma +1}$ are parameters.
\begin{figure*}[t]
\center
\resizebox{0.9\textwidth}{!}{%
  \includegraphics{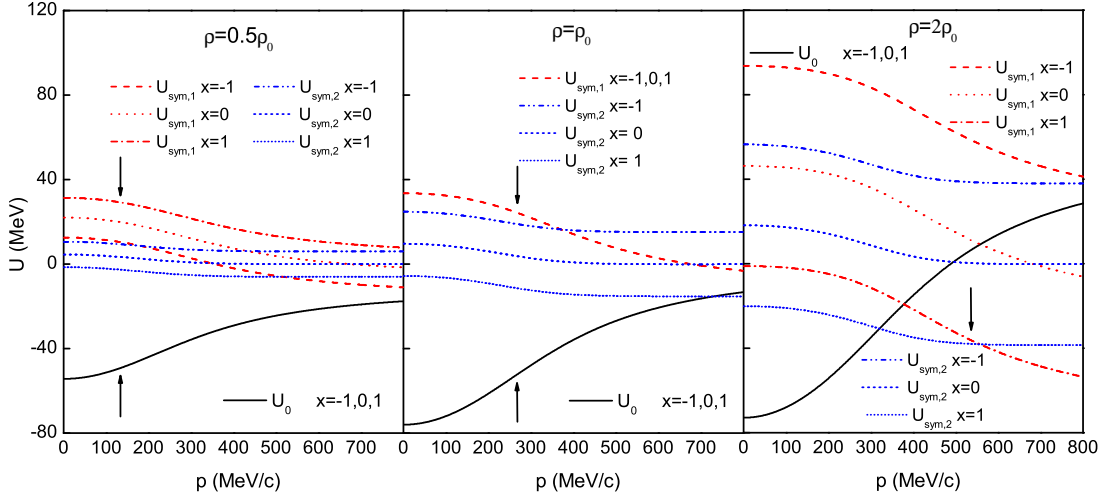}}
\caption{Momentum dependence of the isoscalar and isovector
potentials of the MDI interaction with different x parameters at
densities $0.5\rho_0$ (left panel), $\rho_0$ (middle panel), and
$2\rho_0$ (right panel), respectively. The Fermi momentum $p=p_f$
is denoted by the black arrows.} \label{fig1}
\end{figure*}
By expanding the single-nucleon potential in $\delta$, the isospin
independent isoscalar potential $U_{0}(\rho, p)$ of the MDI
interaction is given by
\begin{eqnarray}
&&U_{0}(\rho, p) = \frac{A_l(x)+A_u(x)}{2}\frac{\rho}{\rho_0} +
B(\frac{\rho}{\rho_0})^{\sigma} + \frac{2(C_l+C_u)}{\rho_0}
\frac{\pi\Lambda^2}{h^3p}
\nonumber \\
&&\times
 \{4p p_f+4 p \Lambda\textrm{ Arctan}[\frac{p-p_f}{\Lambda
}]-4p\Lambda\textrm{ Arctan}[\frac{p+p_f}{\Lambda }]
\nonumber \\
&&-(p^2-p_f^2-\Lambda^2)\textrm{log}[\frac{(p+p_f)^2+\Lambda^2}{(p-P_f)^2+\Lambda^2}]\}.
\end{eqnarray}
The first order symmetry potential of the MDI interaction
$U_{sym,1}(\rho, \vec p)$ is
\begin{eqnarray}
&&U_{sym,1}(\rho, p)=\frac{A_l(x) -
 A_u(x)}{2}\frac{\rho}{\rho_0} - \frac{2x\,B}{\sigma+1} (\frac{\rho}{\rho_0})^{\sigma}
\\ \nonumber
&& + \frac{2(C_l-C_u)}{\rho_0} \frac{2\pi\Lambda^2
p_f^2}{3h^3p}\times
\textrm{log}[\frac{(p+p_f)^2+\Lambda^2}{(p-P_f)^2+\Lambda^2}].
\end{eqnarray}
The second order symmetry potential of the MDI interaction
$U_{sym,2}(\rho,  p)$ is
\begin{eqnarray}
&&U_{sym,2}(\rho, p)=
\\ \nonumber
&&- x\,B (\frac{\rho}{\rho_0})^{\sigma} + \frac{2x\,B}{\sigma+1}
(\frac{\rho}{\rho_0})^{\sigma} + \frac{2(C_l+C_u)}{\rho_0}
\frac{\pi\Lambda^2 p_f^2}{9h^3p}
\\ \nonumber
&&\times
\{\frac{4pp_f(p^2-p_f^2+\Lambda^2)}{[(p-p_f)^2+\Lambda^2][(p+p_f)^2+\Lambda^2]}
\\ \nonumber
&&-\textrm{log}[\frac{(p+p_f)^2+\Lambda^2}{(p-P_f)^2+\Lambda^2}]\}.
\end{eqnarray}
The analytical forms of higher order symmetry potentials
($U_{sym,3}(\rho, p)$...) are not given because only the first
order and second order symmetry potentials ($U_{sym,1}(\rho,  p)$
and $U_{sym,2}(\rho,  p)$  ) are involved in determining the
symmetry energy $E_{sym}(\rho)$ and its density slope $L(\rho)$.

Before we give the detailed results, it is interesting to compare
both the isoscalar and isovector potentials of the MDI interaction
with different x parameters. We show three typical cases of the
MDI interaction with $x=-1$, 0, and 1, respectively.  The
parameter $x$ is very important because it determines the ratio of
contributions of the density-dependent term in the MDI interaction
to the total energy in the isospin singlet channel and triplet
channel. For example, $x$=1 ($x$=-1) means that the
density-dependent term contributes mostly to the T=0 (T=1) channel
\cite{Xuli10a}. Thus, by varying $x$ from -1 to 1, the MDI
interaction covers a large range of uncertainties coming from the
spin(isospin)-dependence of the in-medium many-body forces
\cite{Xuli10a,dec}. The parameter $x$ does not affect the EOS of
symmetric nuclear matter because the $x$ related contributions
from T=0 and T=1 channels can be cancelled out exactly
\cite{Xuli10a}. In Fig.1, the momentum dependence of the
$U_{0}(\rho, p)$, $U_{sym,1}(\rho, p)$, and $U_{sym,2}(\rho, p)$
is plotted at densities $0.5\rho_0$, $\rho_0$, and $2\rho_0$,
respectively. It is clearly seen in Fig.1 that the isoscalar
potential $U_{0}(\rho, p)$ is increasing with the increasing
momentum $p$ while the $U_{sym,1}(\rho, p)$ and $U_{sym,2}(\rho,
p)$ are all decreasing with the increasing momentum $p$. The
$U_{0}(\rho, p)$ does not depend on the parameter x while both the
$U_{sym,1}(\rho,  p)$ and $U_{sym,2}(\rho, p)$ depend closely on
the choice of parameter x. It is seen from Fig.1 that the first
order symmetry potential $U_{sym,1}(\rho, p)$ varies significantly
with different values of x. The only exception is the
$U_{sym,1}(\rho, p)$ at the saturation density $\rho_0$, which is
independent of the choice of parameter x. This is not surprising
because the symmetry energy
$E_{sym}(\rho_0)=\frac{1}{6}\frac{\partial \lbrack
t+U_{0}]}{\partial k}|_{k_{F}^0}\cdot
k_{F}^0+\frac{1}{2}U_{sym,1}(\rho_0 ,k_{F}^0)$ is fixed to be
30.55 MeV with any value of x at $\rho_0$ in the MDI interaction.
Unlike the $U_{sym,1}(\rho, p)$, the second order symmetry
potential $U_{sym,2}(\rho, p)$ is x-parameter dependent at all
densities. Similar to the behavior of the $U_{sym,1}(\rho, p)$,
the $U_{sym,2}(\rho, p)$ is also decreasing with the increasing
momentum $p$ in Fig.1. More importantly, it is found that the
magnitude of $U_{sym,2}(\rho, p)$ becomes comparable to that of
$U_{sym,1}(\rho, p)$ at high momentum ($p\geq 500 $ MeV/c) or
large density ($\rho=2\rho_0$). Although the $U_{sym,2}$ term does
not contribute to the symmetry energy $E_{sym}(\rho_0)$, however,
its contribution to the slope parameter $L(\rho)$ is as large as
$3 U_{sym,2}(\rho,k_F)$ (see Eq.(\ref{Lexp})). Thus the
$U_{sym,2}$ term can not be neglected if its magnitude is
comparable to that of $U_{sym,1}$.
\begin{figure}[b]
\resizebox{0.5\textwidth}{!}{%
  \includegraphics{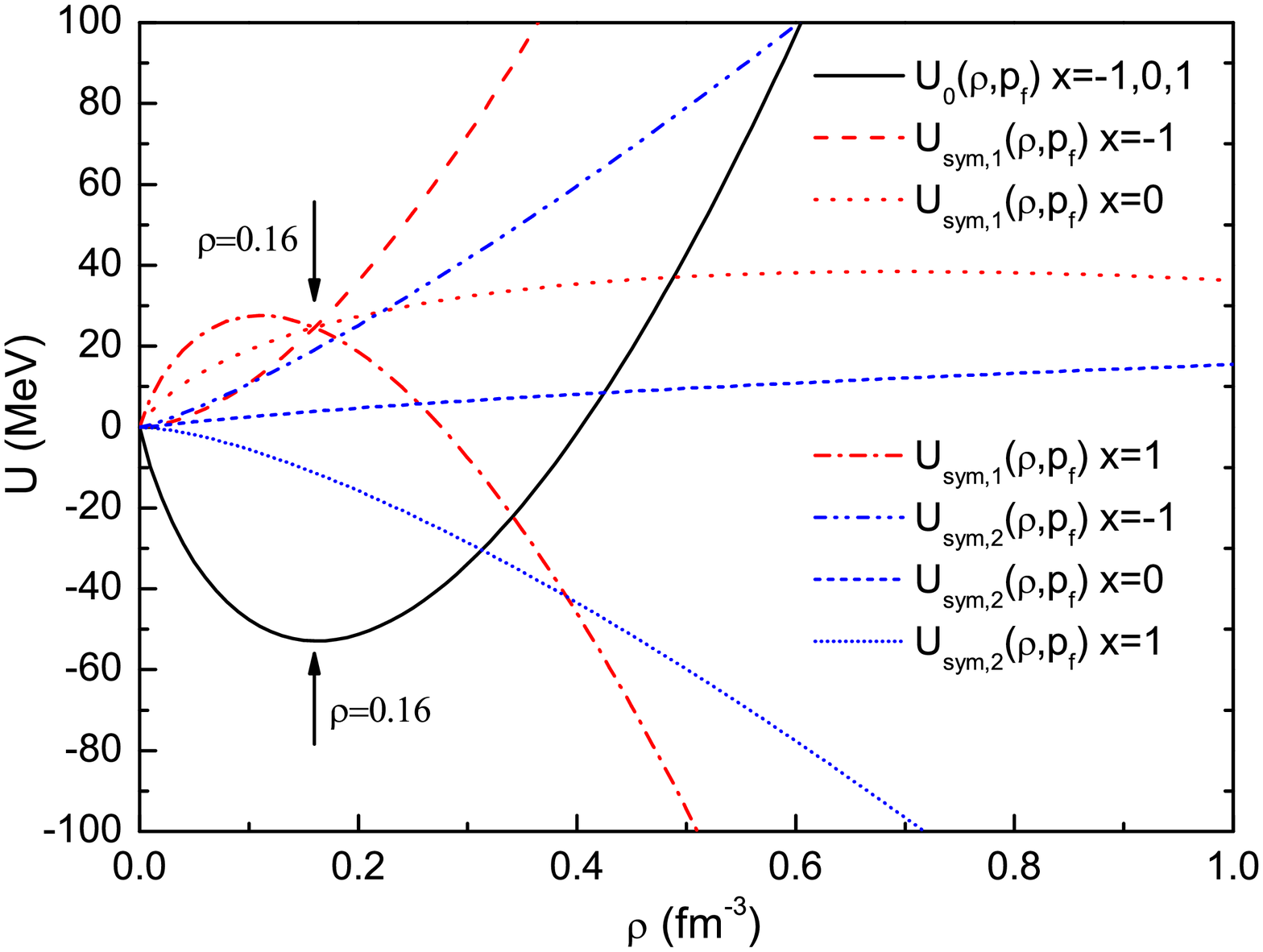}}
\caption{Density dependence of the isoscalar and isovector
potentials of the MDI interaction at $p=p_f$ with x=-1, 0, and 1.
The saturation density $\rho_0$ is denoted by the black arrows.}
\label{fig2}
\end{figure}

In Fig.1, the $U_{0}(\rho, p)$, $U_{sym,1}(\rho, p)$ and
$U_{sym,2}(\rho, p)$ have already shown a strong dependence on the
density $\rho$. To show this density-dependence more clearly, we
plot in Fig.\ref{fig2} the variation of both isoscalar and
isovector potentials with the increasing density at the Fermi
momentum ($p=p_f$). As pointed out in
Ref.\cite{xuli,Das03a,Das03b,Das03c}, it is the $U_{sym,1}(\rho,
p_f)$ in Fig.2 responsible for the rather divergent density
dependence of the symmetry energy $E_{sym}(\rho)$. For instance,
with $x=1$ the $U_{sym,1}$ term decreases very quickly with
increasing density and thus results in a super-soft symmetry
energy at supra saturation densities. On the contrary, the
symmetry energy $E_{sym}(\rho)$ at supra saturation densities is
very stiff for both $x=0$ and $x=-$1 as the contribution of the
$U_{sym,1}$ term becomes very positive with smaller values of $x$.
The $U_{sym,2}(\rho, p)$ term does not contribute to the
$E_{sym}(\rho)$ but do contribute to the density slope
$L({\rho})$. One can see that the $U_{sym,2}$ is actually the most
uncertain part and increases/decreases very fast with the density
when the x parameter equals to -1/1. Thus the magnitude of
$L({\rho})$ is expected to vary significantly with the x-parameter
because of the $U_{sym,2}$ term contribution.
\begin{table*}
\center \caption{Contributions of different terms in the
single-nucleon potential to the symmetry energy $E_{sym}(\rho)$ at
densities $0.5\rho_0$, $1.0\rho_0$, and $2.0\rho_0$.}
\label{tab:1}
\begin{tabular}{ccccccc}
\hline\noalign{\smallskip}
Density & x parameter & kinetic energy  & U$_0$ term   & U$_{sym,1}$ term   & U$_{sym,2}$ term  & Total \\
        &             & contribution    & contribution & contribution       & contribution      &\\
\noalign{\smallskip}\hline\noalign{\smallskip}
                 & -1 & 7.74  & 2.94 & 3.67 & 0 & 14.35 MeV\\
$\rho=0.5\rho_0$ & 0  & 7.74  & 2.94 & 8.38 & 0 & 19.06 MeV\\
                 & 1  & 7.74  & 2.94 & 13.08 & 0 & 23.76 MeV\\
\hline
                 & -1 & 12.29 & 5.96 & 12.30 & 0 & 30.55 MeV\\
$\rho=1.0\rho_0$ & 0  & 12.29 & 5.96 & 12.30 & 0 & 30.55 MeV\\
                 & 1  & 12.29 & 5.96 & 12.30 & 0 & 30.55 MeV\\
\hline
                 & -1 & 19.52 & 11.29 &  40.20 & 0 & 71.01 MeV\\
$\rho=2.0\rho_0$ & 0  & 19.52 & 11.29 &  16.51 & 0 & 47.32 MeV\\
                 & 1  & 19.52 & 11.29 & -7.18  & 0 & 23.63 MeV\\
\noalign{\smallskip}\hline
\end{tabular}
\end{table*}
\begin{table*}
 \center \caption{Contributions of
different terms in the single-nucleon potential to the density
slope $L(\rho)$ at densities $0.5\rho_0$, $1.0\rho_0$, and
$2.0\rho_0$.} \label{tab:2}
\begin{tabular}{ccccccc}
\hline\noalign{\smallskip}
Density & x parameter & kinetic energy  & U$_0$ term   & U$_{sym,1}$ term   & U$_{sym,2}$ term & Total\\
        &             & contribution    & contribution & contribution       & contribution     &  \\
\noalign{\smallskip}\hline\noalign{\smallskip}
                 & -1 & 15.49 & 3.26 & 1.15  & 24.28 & 44.18 MeV\\
$\rho=0.5\rho_0$ & 0  & 15.49 & 3.26 & 15.25 &  6.20 & 40.20 MeV\\
                 & 1  & 15.49 & 3.26 & 29.35 &-11.89 & 36.21 MeV\\
\hline
                 & -1 & 24.59 & 5.69 & 18.34 & 57.22 & 105.84 MeV\\
$\rho=1.0\rho_0$ & 0  & 24.59 & 5.69 & 18.34 & 11.64 & 60.26 MeV\\
                 & 1  & 24.59 & 5.69 & 18.34 &-33.93 & 14.69 MeV\\
\hline
                 & -1 & 39.03 & 9.20 & 87.96 & 135.35 & 271.54 MeV\\
$\rho=2.0\rho_0$ & 0  & 39.03 & 9.20 & 16.88 & 20.49  & 85.60 MeV\\
                 & 1  & 39.03 & 9.20 &-54.20 &-94.35  &-100.32 MeV\\
\noalign{\smallskip}\hline
\end{tabular}
\end{table*}

By using the analytical formulas of Eq.(\ref{Esymexp}) and
Eq.(\ref{Lexp}), the contribution of each term in the
single-nucleon potential to the $E_{sym}(\rho )$ and $L(\rho )$
can be explicitly given. In Table 1 and Table 2, we list the
contributions from the kinetic energy, $U_0$, $U_{sym,1}$ and
$U_{sym,2}$ to the $E_{sym}(\rho)$ and $L(\rho)$ at three
different densities, respectively. As shown in the Tables 1 and 2,
the kinetic energy and the U$_0$ term contributions to the
$E_{sym}(\rho)$ and $L(\rho )$ are the same with different values
of x, which increase smoothly with the increasing density. But the
U$_{sym,1}$ term contribution varies largely with the choice of
x-parameter at abnormal densities, which is clearly the key term
in determining the density dependence of the $E_{sym}(\rho)$ (see
Table 1). For the slope parameter $L(\rho )$, the U$_{sym,2}$ term
becomes as important as the U$_{sym,1}$ term and contributes a
large amount to the $L(\rho )$ (see Table 2). Unlike the kinetic
energy and the U$_0$ term, the contribution of the U$_{sym,1}$ and
U$_{sym,2}$ could be either positive or negative depending on the
choice of the x-parameter.
\begin{figure}[h]
\resizebox{0.48\textwidth}{!}{%
  \includegraphics{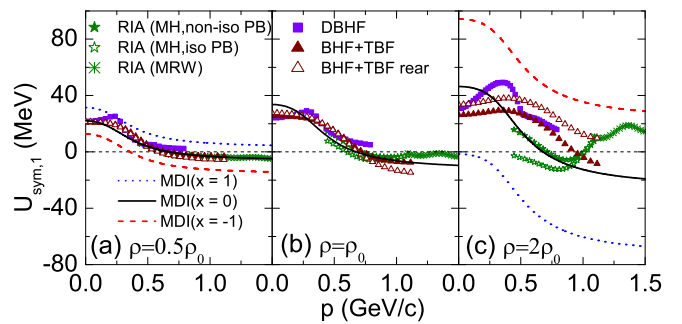}}
\caption{Momentum dependence of the $U_{sym,1}(\protect\rho %
,k)$ at $\protect\rho =0.5\protect\rho _{0}$ (a), $\protect\rho
_{0}$ (b) and $2\protect\rho _{0}$ (c) using the MDI interaction
with $x=-1$, $0$, and $1$. The corresponding results from several
microscopic approaches are also included for comparison (Taken
from Ref.\cite{Che12a}).} \label{Usym1MDI}
\end{figure}
\begin{figure}[h]
\resizebox{0.48\textwidth}{!}{%
  \includegraphics{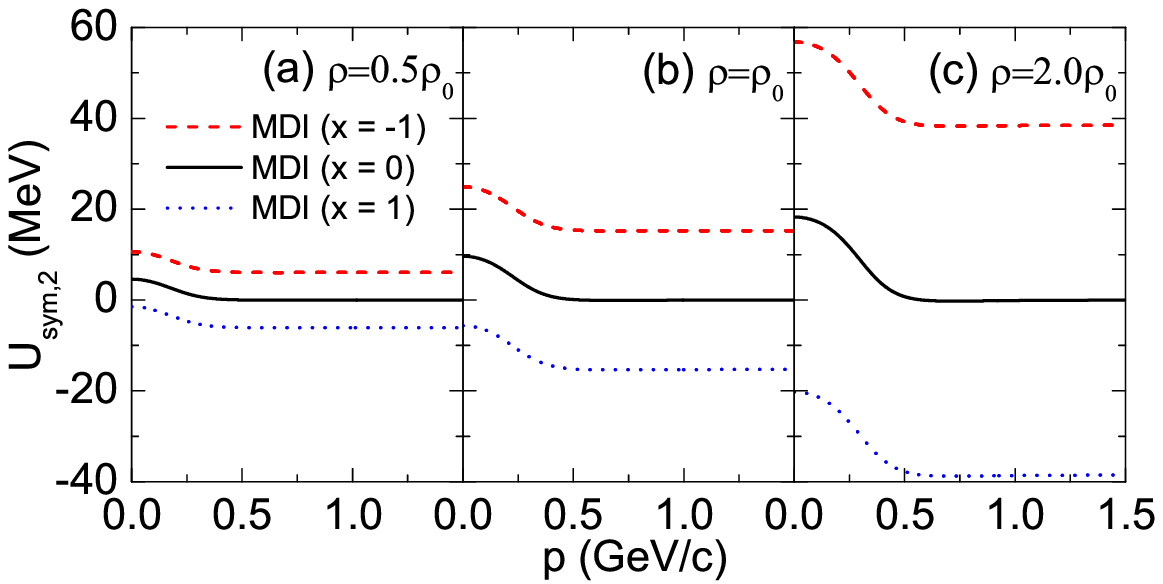}}
  \resizebox{0.48\textwidth}{!}{%
  \includegraphics{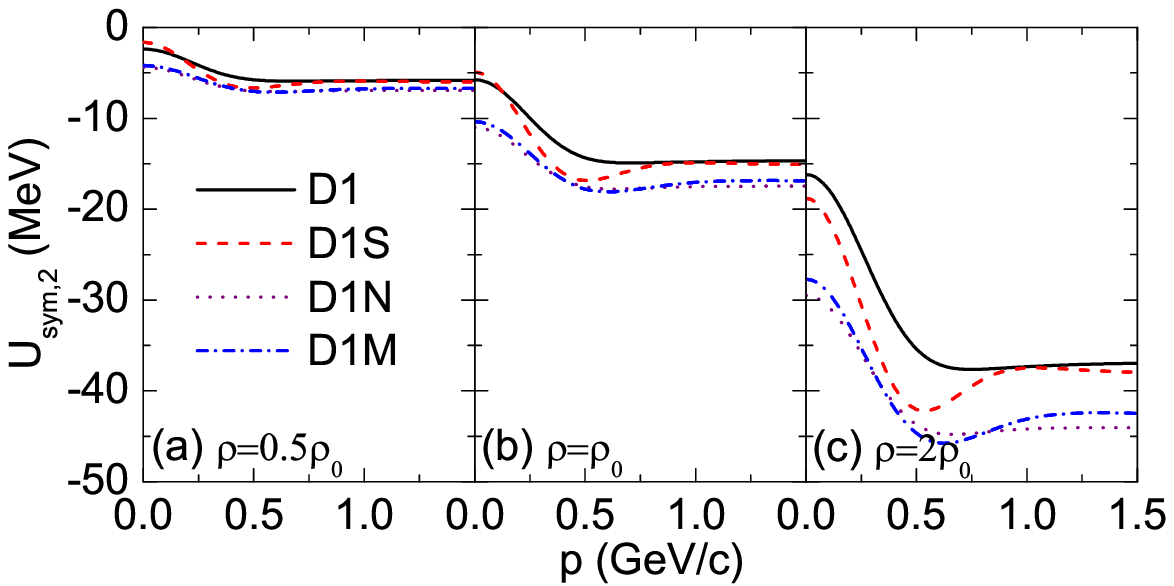}}
\caption{Momentum dependence of the $U_{sym,2}(\protect\rho %
,k)$ at $\protect\rho =0.5\protect\rho _{0}$ (a), $\protect\rho
_{0}$ (b) and $2\protect\rho _{0}$ (c) using the MDI interaction
with $x=-1$, $0$, and $1$ and the Gogny Hartree-Fock\ approach
with D1, D1S, D1N, and D1M (Taken from Ref.\cite{Che12a}).}
\label{Usym2MDI}
\end{figure}

In Fig. \ref{Usym1MDI}, the first order symmetry potential
$U_{sym,1}$ of the MDI interaction is compared with those from
several microscopic approaches, which include the relativistic
impulse approximation (RIA) \cite{Che05c,LiZH06,Mur87,McN83}, the
relativistic Dirac-Brueckner-Hartree-Fock (DBHF) theory
\cite{Dal05}, and the non-relativistic Brueckner-Hartree-Fock
(BHF) theory with/without the 3-body force (TBF) rearrangement
contribution \cite{Zuo06}. For these microscopic results,
it is seen that they are consistent with each other below and around $%
\rho_{0}$. However, there are still larger uncertainties at higher
density of $\rho =2\rho _{0}$. It is interesting to see that the
momentum dependence of the $U_{sym,1}(\rho ,k) $ from the MDI
interaction with $x=0$ agrees well with the results from the
microscopic approaches. As mentioned above, the momentum
dependence of the $U_{sym,1}(\rho ,k)$ at $\rho _{0}$ is the same
for $x=-1$, $0$, and $1$ because $U_{sym,1}(\rho ,k)$ is
independent of the $x$ parameter at saturation density $\rho
_{0}$. In Fig.~\ref{Usym2MDI}, the second-order symmetry potential
$U_{sym,2}$ of the MDI interaction is compared with that from the
Gogny Hartree-Fock approach. From Fig.~\ref{Usym2MDI}, it is
interesting to see that all these interactions firstly decrease
with the momentum and then saturate when the momentum becomes
larger than about ${500}$ MeV/c. Especially the results of the MDI
interaction with ${x=1}$ seem to be in reasonable agreement with
those from the Gogny Hartree-Fock approach.

\section{The optical model analysis of the single-nucleon potential and the corresponding symmetry energy and its density slope}

\begin{figure}[h]
\resizebox{0.45\textwidth}{!}{%
  \includegraphics{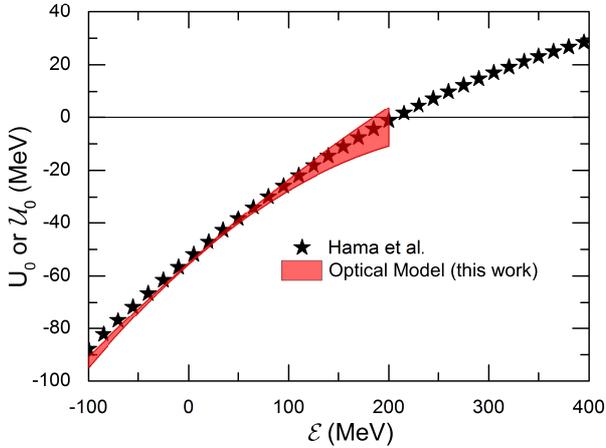}}
\caption{Energy dependence of the isoscalar potential $U_0$ from
the optical model analysis. The results of the
Schr$\ddot{\mathrm{o}}$dinger equivalent potential obtained by
Hama \textit{et al}~\cite{Hama} from the nucleon-nucleus
scattering data are also included for comparison (Taken from
Ref.\cite{Che12b}).} \label{U0}
\end{figure}

The reliable information about the momeumtum/density dependence of
single-nucleon potential (U$_0$, U$_{sym,1}$ and U$_{sym,2}$) in
asymmetric nuclear matter is essential to determine both the
symmetry energy and its density slope. Note that the
momeumtum/density dependence of the isoscalar potential $U_{0}$
around $\rho_0$ has been extensively investigated and relatively
well constrained \cite{pre,BD88}, although there is some
uncertainties at high momeumtum/density. At the saturation
density, the information on $U_{0}$ can be obtained from the
energy dependence of the Global Optical Potential (GOP).
Significant progress has been made in developing the unified GOP
for both nuclear structure and reaction studies over the last
several decades \cite{Hodg,Sat79,Mah91}. The GOP at negative
energies can be constrained by single-nucleon energies of bound
states while at positive energies it is constrained by nuclear
reaction data \cite{Hodg,Sat79,Mah91}. The widely-used expression
of the isoscalar potential is obtained from a large number of
analysis of experimental scattering data and microscopic
calculations \cite{Hodg}
\begin{eqnarray}
U_0(\rho_0,E)=-(50.0-0.30E), \label{U0GOP}
\end{eqnarray}
which gives an effective mass of $ m^*/m=0.7$ and a correct
extrapolation value of $U_0\simeq-54$ MeV at saturation density
($E$=$-16$ MeV), as required by the HVH theorem. Very recently, a
new set of the global isospin dependent neutron-nucleus optical
model potential parameters which include the symmetry potential up
to the second order is obtained for the first time using the
available experimental data from neutron-nucleus scatterings
\cite{Che12b}. Shown in Fig.~\ref{U0} is the energy dependence of
the single-nucleon isoscalar potential $U_0$ obtained in
Ref.\cite{Che12b}. For comparison, the results of the
Schr$\ddot{\mathrm{o}}$dinger equivalent potential obtained by
Hama \textit{et al}~\cite{Hama} from the nucleon-nucleus
scattering data are also shown. It is seen clearly that the
isoscalar potential in Ref.\cite{Che12b} is in good agreement with
that from the Hama's results.

The first order symmetry potential $U_{sym,1}$ can also be deduced
from the optical potential \cite{Hodg,Sat79} using a) elastic
scattering of a neutron and a proton from the same target; b)
proton scattering with the same beam energy on an isotopic chain;
c) (p, n) charge exchange reaction between isobaric analog states.
Since the 1960s, there are several sets of GOPs deduced from
phenomenological model analyses of the available experimental data
\cite{gom1a,gom1b,gom1c,gom1d,gom1e,gom1f,gom2a,gom2b,gom2c,gom2d}.
While some of the analyses assumed an energy independent symmetry
potential, see, \textit{e.g.},
\cite{Hodg,gom2a,gom2b,gom2c,gom2d}, a significant number of
studies considered the energy dependence
\cite{gom1a,gom1b,gom1c,gom1d,gom1e,gom1f}. In these analyses the
symmetry potentials are usually described by using a linear form
$U_{sym,1}(\rho_0,E)=a_{sym}-b_{sym}E$. Assuming that these
various global energy dependent symmetry potentials are equally
accurate with the same predicting power beyond the original energy
ranges in which they were studied, an averaged symmetry potential
\begin{equation}
U_{sym,1}(\rho_0,E)=22.75-0.21E\label{Ubest}
\end{equation}
was obtained \cite{Xuli10b}, which represents the best fit to the
global symmetry potentials constrained by the experimental data up
to date. With this $U_{sym,1}$ and the isoscalar potential $U_0$
in Eq.(\ref{U0GOP}), then the constraints
$E_\mathrm{sym}(\rho_0)=31.3\pm 4.5$ MeV and $L(\rho_0)=52.7\pm
22.5$ MeV were obtained simultaneously by neglecting the
contribution of the second order symmetry potential to $L(\rho_0)$
\cite{Xuli10b}.

\begin{figure}[h]
\resizebox{0.45\textwidth}{!}{%
  \includegraphics{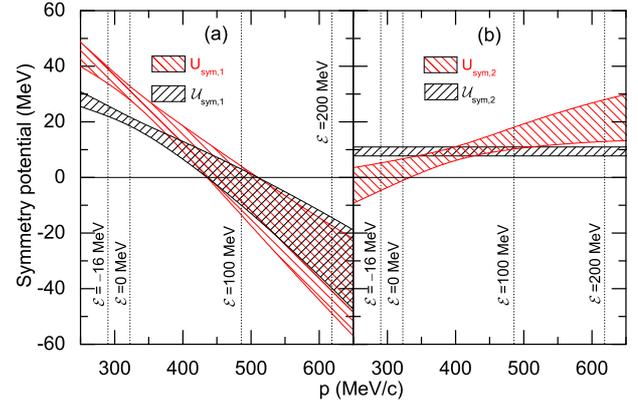}}
\caption{Momentum dependence of the $U_\mathrm{sym,1}$ (a) and
$U_\mathrm{sym,2}$ (b). The corresponding momenta at
$\mathcal{E}=-16$, $0$, $100$ and $200$ MeV are indicated by
dotted lines. The $\mathcal{U}_{\mathrm{sym,1}}$ and
$\mathcal{U}_{\mathrm{sym,2}}$ are terms in the real part of the
central potential of the optical model (Taken from
Ref.\cite{Che12b}).} \label{Usym}
\end{figure}

In contrast to the first order symmetry potential $U_{sym,1}$,
there is very few empirical/experimental information on the second
order symmetry potential $U_{sym,2}$ \cite{Che12a,Che12b}. In the
usual optical model analyses, only the Lane potential $U_{sym,1}$
has been considered. It is thus of great importance to extract
experimental information about the $U_{sym,2}$ and examine its
effects on the density slope of the symmetry energy. Here we show
the information on both the $U_{sym,1}$ and $U_{sym,2}$ from the
very recent optical model analysis of available experimental data
\cite{Che12b}. Shown in Fig.~\ref{Usym} is the momentum dependence
of both the $U_\mathrm{sym,1}$ and $U_\mathrm{sym,2}$ from the
optical model studies. It is seen that the $U_\mathrm{sym,1}$
decreases with momentum $p$ and becomes negative when the momentum
is larger than about $p=470$ MeV/c (i.e., $\mathcal{E}=90$ MeV).
On the contrary, it is interesting to see that the
$U_\mathrm{sym,2}$ increases with the nucleon momentum $p$. It is
also seen that the $U_\mathrm{sym,2}$ essentially vanishes around
$\mathcal{E}=-16$ MeV, as shown in Fig.~\ref{Usym}. Thus the
contribution of the $U_{sym,2}$ term is very small to the density
slope $L(\rho)$ at $\rho_0$, though with a large uncertainty
\cite{Che12b}, verifying the assumption made in
Ref.\cite{Xuli10b}. The new optical model analysis leads to a
value of $E_{\textrm{sym}}(\rho_0)=37.24\pm2.26$ MeV and
$L(\rho_0)=44.98\pm22.31$ MeV, consistent with the results
obtained from analyzing many other observables within various
models.

\section{Summary}
In summary, the general relationship between the symmetry energy
and the single-nucleon potential in isospin-asymmetric matter was
derived by using the Hugenholtz-Van Hove theorem. Both the
symmetry energy $E_{sym}(\rho )$ and its density slope $L(\rho )$
can be expressed explicitly in terms of the magnitude and momentum
dependence of the nucleon isoscalar and isovector potential in
asymmetric nuclear matter. These analytical formulas are useful
for extracting reliable information about the EOS of neutron-rich
nuclear matter from experimental data. Using the isospin and
momentum dependent MDI interaction model as an example, the
contributions of different terms in the single-nucleon potential
(MDI) to the $E_{sym}(\rho )$ and $L(\rho )$ are analyzed in
detail for different densities. The first-order symmetry potential
is found to be responsible for the uncertain high density behavior
of the $E_{sym}(\rho )$ while the density slope $L(\rho )$ depend
on both the first-order and second-order symmetry potentials. By
using the derived analytical formulas and the single-nucleon
potentials from the optical model analysis, both the symmetry
energy $E_{sym}(\rho )$ and its density slope $L(\rho )$ at the
saturation density $\rho_0$ were extracted. To further constrain
the $L(\rho )$ at high densities, more reliable information about
the second-order symmetry potential $U_\mathrm{sym,2}$ is useful.
\\
\\
\section*{Acknowledgments}
The authors would like to thank Bao-Jun Cai, Rong Chen, and
Xiao-Hua Li for fruitful collaboration and stimulating
discussions. This work is supported by the National Natural
Science Foundation of China (Grant Nos 11175085, 11235001,
11035001, 11135011, and 11275125), by the Project Funded by the
Priority Academic Program Development of Jiangsu Higher Education
Institutions (PAPD), by the Shanghai Rising-Star Program under
grant No. 11QH1401100, the ``Shu Guang" project supported by
Shanghai Municipal Education Commission and Shanghai Education
Development Foundation, the Program for Professor of Special
Appointment (Eastern Scholar) at Shanghai Institutions of Higher
Learning, the Science and Technology Commission of Shanghai
Municipality (11DZ2260700), and by the US National Aeronautics and
Space Administration under grant NNX11AC41G issued through the
Science Mission Directorate, the US National Science Foundation
under Grant No. PHY-1068022 and the CUSTIPEN (China-U.S. Theory
Institute for Physics with Exotic Nuclei) under DOE grant number
DE-FG02-13ER42025.

\end{document}